\documentclass[12pt]{article}
\usepackage{graphicx}
\usepackage{epstopdf}
\usepackage{epsfig}
\usepackage{amsmath}
\usepackage{multirow}
\usepackage{subfigure}
\usepackage{rotating}
\usepackage{array}
\def\lsim{\mathrel{\raise.3ex\hbox{$<$\kern-.75em\lower 1ex\hbox{$\sim$}}}}
\def\gsim{\mathrel{\raise.3ex\hbox{$>$\kern-.75em\lower 1ex\hbox{$\sim$}}}}

\def\be{\begin{equation}}
\def\ee{\end{equation}}
\def\bea{\begin{eqnarray*}}
\def\eea{\end{eqnarray*}}

\begin{document}
\title{Seesaw mechanism with four texture zeros \\
in the neutrino Yukawa matrix}
\author{Jiajun Liao$^1$, D. Marfatia$^2$, and K. Whisnant$^1$\\
\\
\small\it $^1$Department of Physics and Astronomy, Iowa State University, Ames, IA 50011, USA\\
\small\it $^2$Department of Physics and Astronomy, University of Kansas, Lawrence, KS 66045, USA}
\date{}
\maketitle

\begin{abstract}

With the recent observation of nonzero $\theta_{13}$, five neutrino
oscillation parameters are now known. By imposing four zeros in the
Yukawa coupling matrix of the type I seesaw model, the number of
parameters in the neutrino mass matrix is reduced to seven, and
we are able to make predictions for the lightest neutrino
mass, Dirac CP phase, and neutrinoless double
beta decay. Four texture zeros in the Yukawa coupling
matrix is equivalent to either a single texture zero or a single
cofactor zero for an off-diagonal element of the light neutrino mass
matrix. We find strong similarities between single texture
zero models with one mass ordering and single cofactor zero models
with the opposite mass ordering. In the context of a specific class
of single-flavor leptogenesis models, we find additional constraints
on the parameter space.

\end{abstract}

\newpage

\section{Introduction}

Low-energy neutrino phenomenology is described by nine parameters
in the Majorana mass matrix of light neutrinos, which can be written
as~\cite{PDG}
$M=V^*\text{diag}(m_1, m_2, m_3)V^\dagger$, with $V = U\text{diag}(1,
e^{i\phi_2/2}, e^{i\phi_3/2})$ and
\begin{align}
U=\begin{bmatrix}
   c_{13}c_{12} & c_{13}s_{12} & s_{13}e^{-i\delta} \\
   -s_{12}c_{23}-c_{12}s_{23}s_{13}e^{i\delta} & c_{12}c_{23}-s_{12}s_{23}s_{13}e^{i\delta} & s_{23}c_{13} \\
   s_{12}s_{23}-c_{12}c_{23}s_{13}e^{i\delta} & -c_{12}s_{23}-s_{12}c_{23}s_{13}e^{i\delta} & c_{23}c_{13}
   \end{bmatrix}.
\label{eq:U}
\end{align}
After the measurement of $\theta_{13}$ by the Daya
Bay~\cite{An:2012eh}, RENO~\cite{Ahn:2012nd}, and Double
Chooz~\cite{DC} experiments, five of them are known; the result of a
recent global three-neutrino fit~\cite{Fogli:2012ua} is shown in
Table~1.  The tiny masses of light neutrinos can be elegantly
explained by the seesaw mechanism~\cite{seesaw}, but, unfortunately,
with the introduction of additional free parameters that cannot be
measured in the forseeable future. The most popular seesaw model is
the type I seesaw, in which the mass matrix of the light neutrinos can
be written as
\begin{align}
M=\lambda^T {M_R}^{-1}\lambda v^2=Y^TY\,,
\label{eq:Mseesaw}
\end{align}
where, for $N$ heavy right-handed neutrinos, ${M_R}=\text{diag}(M_1, M_2, \cdots, M_N)$,  $\lambda$ is a $N\times 3$ Yukawa coupling matrix, and
\begin{equation}
Y=v {M_R}^{-1/2}\lambda=\begin{bmatrix}
      y_{1e} & y_{1\mu} & y_{1\tau} \\
      y_{2e} & y_{2\mu} & y_{2\tau} \\
      \vdots & \vdots   & \vdots     \\
      y_{Ne} & y_{N\mu} & y_{N\tau}
      \end{bmatrix}.
\label{eq:Ymatrix}
\end{equation}
Note that on permuting the rows of the Y matrix (which is equivalent
to reordering the right-handed neutrinos) or applying a rotation to the
rows of the $Y$ matrix (which is equivalent to a rotation in the space
of the right-handed neutrinos), the mass matrix of the light neutrinos
remains the same.

The standard seesaw model has $N=3$ and permits any set of low-energy
neutrino parameters in $M_\nu$. A way to extract predictions from the
seesaw model is to impose constraints on its parameters. The most
economical seesaw model includes two right-handed neutrinos and two
zeros in the Yukawa coupling matrix, or, equivalently, two zeros in
$Y$~\cite{Frampton:2002qc}. 
The measurement of $\theta_{13}$ excludes the normal mass ordering (NO)
and stringently constrains the allowed parameter space for the
inverted mass ordering (IO); in particular, the phase
$\delta$ must be such that Dirac CP violation is close to
maximal~\cite{Harigaya:2012bw}.
Also, for two right-handed neutrinos, one of the light
neutrinos must have vanishing mass, so the allowed values of the sum of
all light neutrino masses (which affects structure formation in our universe) and $|M_{ee}|$ (which determines the rate for neutrinoless double-beta decay ($0\nu\beta\beta$), a signal
of lepton number violation) are limited.

\begin{table}
\caption{Best-fit values and $2\sigma$ ranges of the oscillation
parameters~\cite{Fogli:2012ua}, with $\delta m^2 \equiv m_2^2-m_1^2$
and $\Delta m^2 \equiv m_3^2-(m_1^2+m_2^2)/2$.
\label{tab:data}}
\begin{center}
\begin{tabular}{|l|*{5}{c|}}\hline
\makebox[4em]{Ordering}
&\makebox[2em]{$\theta_{12}(^\circ)$}&\makebox[2em]{$\theta_{13}(^\circ)$}&\makebox[2em]{$\theta_{23}(^\circ)$}
&\makebox[6em]{$\delta m^2(10^{-5}\text{eV}^2)$}&\makebox[7em]{$|\Delta m^2|(10^{-3}\text{eV}^2)$}\\\hline
Normal &$33.6^{+2.1}_{-2.0}$&$8.9^{+0.9}_{-0.9}$&$38.4^{+3.6}_{-2.3}$&$7.54^{+0.46}_{-0.39}$&$2.43^{+0.12}_{-0.16}$\\\hline
Inverted &$33.6^{+2.1}_{-2.0}$&$9.0^{+0.8}_{-1.0}$&$38.8^{+5.3}_{-2.3}\oplus 47.5-53.2$&$7.54^{+0.46}_{-0.39}$&$2.42^{+0.11}_{-0.16}$\\\hline
\end{tabular}
\end{center}
\end{table}

In this paper, we extend this most economical model to include a third
right-handed neutrino. We use texture zeros in $Y$ as extra constraints,
which may arise in, e.g., extra dimensional
models~\cite{Harigaya:2012bw}. If there are 5 or more texture zeros in
$Y$, the most economical model with two right-handed neutrinos
or a block diagonal matrix is obtained; the latter
is excluded by the current experimental data.
So the simplest case
for three right-handed neutrinos has four texture zeros in $Y$,
which is equivalent to five nonzero elements; for previous work see
Refs.~\cite{Branco:2007nb,Randhawa:2002jt}.
Here we derive analytic formulas that relate the free parameters to the
dependent ones and determine the constraints on these models including
the recent data on $\theta_{13}$.

With five nonzero elements in $Y$ only two phases are physical, so
that there are seven free parameters in the four texture zero model.
Hence, we can use the five observed oscillation parameters from the
global fit to determine the allowed regions for the Dirac CP phase
$\delta$ and $m_1$ ($m_3$) for the normal (inverted) ordering. Then
we can obtain the values of the Majorana phases $\phi_2$ and $\phi_3$,
completing our knowledge of all the elements in the light neutrino
mass matrix. Furthermore, from Eq.~(\ref{eq:Mseesaw})
we can find $Y$ and then study neutrinoless double beta decay and
leptogenesis.

In Sec. 2, we list the general properties of the four texture zero
model and give a brief review of neutrinoless double beta decay
and leptogenesis. In Sec. 3, we use current experimental data to study
the allowed parameter regions for each texture. We conclude in Sec.~4.

\section{Classes with four texture zeros}

There are 3 basic ways to have 5 nonzero elements: (2, 2, 1), (3, 1,
1) or (3, 2, 0), where the numbers indicate how many nonzero elements
are in each row of the $Y$ matrix. The (3, 2, 0) case is equivalent to
two right-handed neutrinos with one texture zero element and one vanishing
mass, and is equivalent to the most economical $N=2$ model after
a rotation in the right-handed neutrino space. So only the (2, 2, 1)
and (3, 1, 1) cases need to be considered. There are 9 independent
textures in the (2, 2, 1) case and 3 independent textures in the (3,
1, 1) case. We divide them into four classes:

Class 1\begin{align}
   1A:\begin{bmatrix}
   \times & \times & 0 \\
   \times & 0 & \times  \\
   \times & 0 & 0
   \end{bmatrix}\quad
   1B:\begin{bmatrix}
   \times & \times & 0 \\
   \times & 0 & \times  \\
   0 & \times & 0
   \end{bmatrix}\quad
   1C:\begin{bmatrix}
   \times & \times & 0 \\
   \times & 0 & \times  \\
   0 & 0 & \times
   \end{bmatrix}\nonumber
\end{align}

Class 2\begin{align}
   2A:\begin{bmatrix}
   \times & \times & 0 \\
   0 & \times & \times  \\
   \times & 0 & 0
   \end{bmatrix}\quad
   2B:\begin{bmatrix}
   \times & \times & 0 \\
   0 & \times & \times  \\
   0 & \times & 0
   \end{bmatrix}\quad
   2C:\begin{bmatrix}
   \times & \times & 0 \\
   0 & \times & \times  \\
   0 & 0 & \times
   \end{bmatrix}\nonumber
\end{align}

Class 3\begin{align}
   3A:\begin{bmatrix}
   \times & 0 & \times \\
   0 & \times & \times  \\
   \times & 0 & 0
   \end{bmatrix}\quad
   3B:\begin{bmatrix}
   \times & 0 & \times \\
   0 & \times & \times  \\
   0 & \times & 0
   \end{bmatrix}\quad
   3C:\begin{bmatrix}
   \times & 0 & \times \\
   0 & \times & \times  \\
   0 & 0 & \times
   \end{bmatrix}\nonumber
\end{align}

Class 4\begin{align}
   4A:\begin{bmatrix}
   \times & \times & \times \\
   0 & \times & 0  \\
   0 & 0 & \times
   \end{bmatrix}\quad
   4B:\begin{bmatrix}
   \times & \times & \times \\
   0 & 0 & \times  \\
   \times & 0 & 0
   \end{bmatrix}\quad
   4C:\begin{bmatrix}
   \times & \times & \times \\
   \times & 0 & 0  \\
   0 & \times & 0
   \end{bmatrix}\nonumber
\end{align}
For each class there are six possible permutations of the rows, so
there are 72 individual cases in all~\cite{Branco:2007nb}.

The rate for neutrinoless double beta decay depends on the effective
Majorana mass, which is equal to the magnitude of the $\nu_e-\nu_e$
element of the neutrino mass matrix,
\be
|M_{ee}|=|m_1c_{12}^2c_{13}^2+m_2 e^{-i\phi_2}s_{12}^2c_{13}^2+m_3 e^{-i\phi_3}s_{13}^2e^{2i\delta}|.
\ee
Note that $|M_{ee}|$ can be written in a form that is independent of the Dirac CP phase $\delta$ by redefining $\phi_3$. We choose the form above because the Majorana phases will be constrained. The latest experimental result from EXO-200~\cite{Auger:2012ar} shows that
the effective mass is less than $140-380$ meV at 90\% C.L. In the
foreseeable future, experiments such as KamLAND-Zen will reach a
sensitivity of about 50 meV or below~\cite{Rodejohann:2012xd}.

In principle, leptogenesis may provide hints of the 
structure of the Yukawa matrix; see Ref.~\cite{Blanchet:2012bk}
for a recent review.
In what follows, we adopt a minimal scenario of
leptogenesis~\cite{Giudice:2003jh}.  We work in the single flavor
approximation in which $M_1 > 10^{12}$~GeV, so that the flavor
composition of the leptons does not affect the baryon asymmetry of
the universe. We also assume the right-handed neutrinos to be hierarchical,
$M_2,M_3\gg M_1>10^{12}$ GeV. Then in the standard model the baryon
asymmetry is given by~\cite{Giudice:2003jh}
\be
\eta_{B0}=\frac{n_B}{n_\gamma}=-9.72\times 10^{-3}\times\epsilon_1\times \eta,
\ee
where $\epsilon_1$ is the CP asymmetry in the decay of the lightest
right-handed neutrino, and $\eta$ is the wash out efficiency factor, which
can be obtained by solving the Boltzmann equation. We use
a very simple analytic fit~\cite{Giudice:2003jh}
\be
\eta \simeq \left(\frac{0.55\times10^{-3}\text{ eV}}{\tilde{m}_1}\right)^{1.16},
\label{eq:eta}
\ee
where $\tilde{m}_1$ is
\be
\tilde{m}_1=\sum_{\substack{\alpha=e,\mu,\tau}}|\lambda_{1\alpha}|^2\frac{v^2}{M_1}=\sum_{\substack{\alpha=e,\mu,\tau}}|y_{1\alpha}|^2.
\label{eq:meff1}
\ee
Equation~(\ref{eq:eta}) is valid for $M_1\ll 10^{14}$ GeV and
$\tilde{m}_1\geq0.01$ eV. With these constraints, the baryon asymmetry is
\be
\eta_{B0}=\frac{n_B}{n_\gamma}\simeq-3.4\times10^{-4}\times\epsilon_1(\frac{0.01\text{eV}}{\tilde{m}_1})^{1.16},
\label{eq:etaB}
\ee
where $\epsilon_1$  can be written as 
\begin{align}
\epsilon_1&=-\frac{3}{16\pi}\frac{M_1}{(\lambda\lambda^\dagger)_{11}}\text{Im}[(\lambda\lambda^\dagger M_R^{-1}(\lambda\lambda^\dagger)^T)_{11}]\nonumber \\
&=-\frac{3}{16\pi}\frac{1}{(\lambda\lambda^\dagger)_{11}}\sum_{\substack{j\neq 1}}\text{Im}(\lambda\lambda^\dagger)_{1j}^2\frac{M_1}{M_j}.
\end{align}
Since  $\lambda=\frac{1}{v} M_R^{1/2}Y$, we have
\begin{align}
\epsilon_1&=-\frac{3}{16\pi}\frac{M_1}{v^2(YY^\dagger)_{11}}\sum_{\substack{j\neq 1}}\text{Im }(YY^\dagger)_{1j}^2\nonumber \\
&=-\frac{3M_1}{16\pi v^2}\sum_{\substack{j\neq 1}}\frac{\text{Im}(y_{1e}y_{je}^*+y_{1\mu}y_{j\mu}^*+y_{1\tau}y_{j\tau}^*)^2}{|y_{1e}|^2+|y_{1\mu}|^2+|y_{1\tau}|^2}.
\label{eq:epsilon1}
\end{align}

Hence, the baryon asymmetry is proportional to the lightest right
handed neutrino mass $M_1$. The observed baryon asymmetry is
$\eta_{B0}=6.19\times10^{-10}$~\cite{Komatsu:2010fb}. Since $Y$
is determined by the light neutrino mass matrix, we can
calculate the lightest right-handed neutrino mass $M_1$ from
Eqs.~(\ref{eq:meff1}), (\ref{eq:etaB}), and (\ref{eq:epsilon1}), ensuring
that $10^{12} \text{ GeV}<M_1\ll
10^{14}$ GeV.

\section{Phenomenology}

\subsection{Class 1 ($M_{23} = 0$)}

The mass matrices of the three textures in Class 1 always
have a zero in the (2, 3) entry. In fact, \(M_{23}=0\) is the only
condition on $Y$ for all the textures in Class 1, and the
condition is the same for both mass orderings (normal and inverted). Take Class 1A for example; we can write
\begin{align}
Y=\begin{bmatrix}
   a & b & 0 \\
   c & 0 & d  \\
   e & 0 & 0
   \end{bmatrix}\,,
\end{align}
where $a,b,c,d,e$ are all nonzero complex numbers. Then the mass matrix of the light neutrinos becomes
\begin{align}
M=Y^TY=\begin{bmatrix}
   a^2+c^2+e^2 & ab & cd \\
   ab & b^2 & 0  \\
   cd & 0 & d^2
   \end{bmatrix}.
\label{eq:M1A}
\end{align}
Comparing Eq.~(\ref{eq:M1A}) with the standard parametrization, if
\(M_{23}=0\), we can find a solution for \(Y\) as follows:
\begin{align}
b=\pm M_{22}^{1/2},a=M_{12}/b,d=\pm M_{33}^{1/2},c=M_{13}/d,e=\pm (M_{11}-a^2-c^2)^{1/2}.
\label{eq:solution}
\end{align}
Since \(a,b,c,d,e\) are all nonzero complex numbers, a solution
always exists. Solutions for $Y$ for the other two textures
in Class 1 may be derived in a similar fashion. Therefore, Class 1 is defined
by the necessary and sufficient condition $M_{23} = 0$, which
can be written as
\be
m_1 = - {m_3 e^{i\phi_3} U_{\tau3}U_{\mu3}+m_2 e^{i\phi_2} U_{\tau2}U_{\mu2}\over U_{\tau1}U_{\mu1}}.
\label{eq:condition1}
\ee
Taking the absolute square gives
\be
m_1^2|U_{\mu1}|^2|U_{\tau1}|^2
-m_2^2|U_{\mu2}|^2|U_{\tau2}|^2
-m_3^2|U_{\mu3}|^2|U_{\tau3}|^2
= 2{\rm Re}(m_3 e^{-i\phi_3} U_{\mu3}^* U_{\tau3}^* m_2 e^{i\phi_2} U_{\mu2} U_{\tau2}),
\ee
or, defining $\phi = \phi_3-\phi_2$, 
\begin{eqnarray}
&m_1^2&|U_{\mu1}|^2|U_{\tau1}|^2
-m_2^2|U_{\mu2}|^2|U_{\tau2}|^2
-m_3^2|U_{\mu3}|^2|U_{\tau3}|^2
\nonumber\\
&=& -2 m_2 m_3 c_{13}^2 s_{23} c_{23}
{\rm Re}\left[e^{-i\phi} (c_{12}c_{23} - s_{12} s_{23} s_{13} e^{i\delta})
(c_{12} s_{23} + s_{12} c_{23} s_{13} e^{i\delta}) \right]
\nonumber\\
&=& -2 m_2 m_3 c_{13}^2 s_{23} c_{23} \times
\nonumber\\
&~& {\rm Re}\left[c_{12}^2 s_{23} c_{23} \cos\phi
+ s_{12} c_{12} s_{13} (c_{23}^2-s_{23}^2) \cos(\phi-\delta)
- s_{12}^2 s_{23} c_{23} s_{13}^2 \cos(\phi-2\delta) \right].
\end{eqnarray}
Expanding the cosines yields the form
\begin{align}
C = A\cos\phi + B\sin\phi,
\label{eq:condition}
\end{align}
with A, B and C as listed in Table~\ref{tab:ABC}.

\begin{sidewaystable}
\centering
$C=A\cos\phi + B\sin\phi$\\
\begin{tabular}{|c|c|c|c|}\hline
Class & A & B & C \\
\hline
1&$\begin{array}{l}
-2m_2m_3 c_{13}^2 s_{23} c_{23}\times\\
\left[c_{12}^2 s_{23} c_{23}+ s_{12} c_{12} s_{13} (c_{23}^2-s_{23}^2) \cos\delta\right.\\
\left.- s_{12}^2 s_{23} c_{23} s_{13}^2 \cos(2\delta)\right]
\end{array}$&$\begin{array}{l} 
-2m_2m_3 c_{13}^2 s_{23} c_{23} s_{12} s_{13}\times\\
\left[c_{12} (c_{23}^2-s_{23}^2) \sin\delta\right.\\
\left.- s_{12} s_{23} c_{23} s_{13} \sin(2\delta)\right]
\end{array}$&$\begin{array}{l}
m_1^2|U_{\mu1}|^2|U_{\tau1}|^2\\
-m_2^2|U_{\mu2}|^2|U_{\tau2}|^2\\
-m_3^2|U_{\mu3}|^2|U_{\tau3}|^2
\end{array}$\\
\hline
2&$\begin{array}{l}
-2m_2m_3 c_{13}^2 s_{12} c_{23} s_{13} \times\\
\left[c_{12} s_{23} \cos\delta + s_{12} c_{23} s_{13} \cos(2\delta)\right]
\end{array}$&$\begin{array}{l} 
-2m_2m_3 c_{13}^2 s_{12} c_{23} s_{13}\times\\
\left[c_{12} s_{23} \sin\delta + s_{12} c_{23} s_{13} \sin(2\delta) \right]
\end{array}$&$\begin{array}{l}
 m_1^2|U_{e1}|^2|U_{\tau1}|^2\\
-m_2^2|U_{e2}|^2|U_{\tau2}|^2\\
-m_3^2|U_{e3}|^2|U_{\tau3}|^2
\end{array}$\\
\hline
3&$\begin{array}{l}
2m_2m_3 c_{13}^2 s_{12} s_{23} s_{13} \times\\
\left[c_{12} c_{23} \cos\delta - s_{12} s_{23} s_{13} \cos(2\delta) \right]
\end{array}$&$\begin{array}{l} 
2m_2m_3 c_{13}^2 s_{12} s_{23} s_{13} \times\\
\left[c_{12} c_{23} \sin\delta - s_{12} s_{23} s_{13} \sin(2\delta) \right]
\end{array}$&$\begin{array}{l}
m_1^2|U_{e1}|^2|U_{\mu1}|^2\\
-m_2^2|U_{e2}|^2|U_{\mu2}|^2\\
-m_3^2|U_{e3}|^2|U_{\mu3}|^2
\end{array}$\\
\hline
4A&$\begin{array}{l}
-2m_2^{-1}m_3^{-1} c_{13}^2s_{23}c_{23}\times\\
\left[c_{12}^2 s_{23} c_{23}+ s_{12} c_{12} s_{13}(c_{23}^2-s_{23}^2) \cos\delta\right.\\
\left.- s_{12}^2 s_{23} c_{23} s_{13}^2 \cos(2\delta)\right]
\end{array}$&$\begin{array}{l} 
 -2m_2^{-1}m_3^{-1} c_{13}^2s_{23}c_{23}s_{12}s_{13}\times\\
\left[c_{12}(c_{23}^2-s_{23}^2) \sin\delta\right.\\
\left.- s_{12} s_{23} c_{23} s_{13} \sin(2\delta) \right]
\end{array}$&$\begin{array}{l}
m_1^{-2}|U_{\mu1}|^2|U_{\tau1}|^2\\
- m_2^{-2}|U_{\mu2}|^2|U_{\tau2}|^2\\
- m_3^{-2}|U_{\mu3}|^2|U_{\tau3}|^2
\end{array}$\\
\hline
4B&$\begin{array}{l}
-2m_2^{-1}m_3^{-1} c_{13}^2 s_{12} c_{23} s_{13} \times\\
\left[c_{12} s_{23} \cos\delta + s_{12} c_{23} s_{13} \cos(2\delta) \right]
\end{array}$&$\begin{array}{l} 
-2m_2^{-1}m_3^{-1} c_{13}^2 s_{12} c_{23} s_{13} \times\\
\left[c_{12} s_{23} \sin\delta + s_{12} c_{23} s_{13} \sin(2\delta) \right]
\end{array}$&$\begin{array}{l}
m_1^{-2}|U_{e1}|^2|U_{\tau1}|^2\\
- m_2^{-2}|U_{e2}|^2|U_{\tau2}|^2\\
- m_3^{-2}|U_{e3}|^2|U_{\tau3}|^2
\end{array}$\\
\hline
4C&$\begin{array}{l}
2m_2^{-1}m_3^{-1} c_{13}^2 s_{12} c_{23} s_{13} \times\\
\left[c_{12} c_{23} \cos\delta - s_{12} s_{23} s_{13} \cos(2\delta) \right]
\end{array}$&$\begin{array}{l} 
2m_2^{-1}m_3^{-1} c_{13}^2 s_{12} c_{23} s_{13} \times\\
\left[c_{12} c_{23} \sin\delta - s_{12} s_{23} s_{13} \sin(2\delta) \right]
\end{array}$&$\begin{array}{l}
m_1^{-2}|U_{e1}|^2|U_{\mu 1}|^2\\
- m_2^{-2}|U_{e2}|^2|U_{\mu 2}|^2\\
- m_3^{-2}|U_{e3}|^2|U_{\mu 3}|^2
\end{array}$\\
\hline
\end{tabular}
\caption{The coefficients A, B and C for each class.
\label{tab:ABC}}
\end{sidewaystable}

\begin{figure}
\centering
\includegraphics[width=6.0in]
{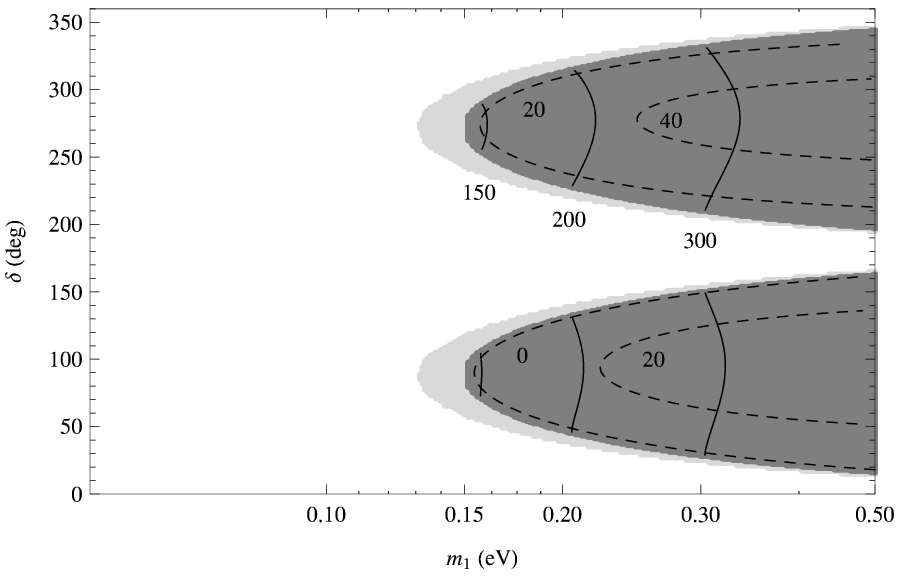}
\caption{The allowed regions in the $(m_1,\delta)$ plane for Class 1 and
the normal ordering. The dark shaded regions correspond to the best-fit
parameters of the oscillation parameters, while the light shaded regions
are allowed at $2\sigma$. The solid lines are iso-$|M_{ee}|$ contours
(in meV) and the dashed lines are iso-$\phi_2$ contours (in degrees).}
\label{fg:1ANH}
\end{figure}

\begin{figure}
\centering
\includegraphics[width=6.0in]
{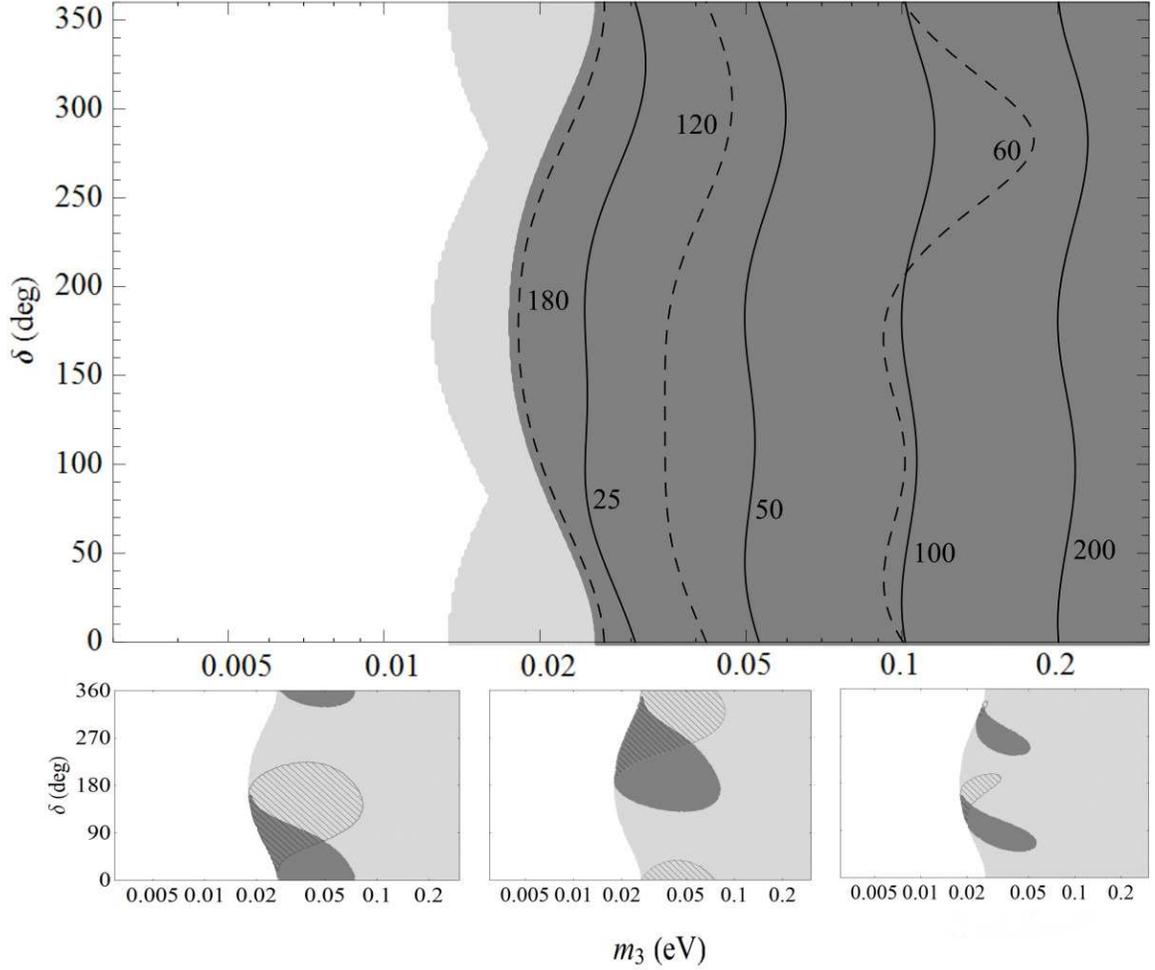}

\caption{The upper panel shows the allowed regions in the $(m_3,\delta)$
plane for Class 1 and the inverted ordering. The shading and line types
in the upper panel are as in Fig.~1. The lower panels show the allowed
regions for Class 1A and the inverted ordering with the additional
constraint of successful single-flavored leptogenesis. The hatched (dark
shaded) regions use the plus (minus) solution of $\phi_2$. From left
to right the three graphs have the first, second and third row of
$Y$ associated with the lightest right-handed neutrino mass,
respectively.}
\label{fg:1AIH}
\end{figure}

For Eq.~(\ref{eq:condition}), if $C^2 > A^2 + B^2$, there is no
solution; if $C^2 < A^2 + B^2$, there are two solutions:
\begin{equation}
\phi=2\arctan\frac{B \pm \sqrt{A^2+B^2-C^2}}{A+C}.
\label{eq:phi}
\end{equation}
We can write Eq.~(\ref{eq:condition1}) in terms of $\phi$ as
\be
m_1 =e^{i\phi_2}\frac{-m_3 e^{i\phi}U_{\tau3}U_{\mu 3}-m_2 U_{\tau2}U_{\mu 2}} {U_{\tau1}U_{\mu 1}}.
\ee
Since $m_1$ is a non-negative real number, we get
\begin{equation}
\phi_2=-\text{arg}[\frac{-m_3 e^{i\phi}U_{\tau3}U_{\mu 3}- m_2 U_{\tau2}U_{\mu 2}} {U_{\tau1}U_{\mu 1}}],
\label{eq:phi2}
\end{equation}
and
\begin{equation}
\phi_3=\phi_2+\phi.
\label{eq:phi3}
\end{equation}
Therefore, if $m_1$ ($m_3$) and $\delta$ in the normal (inverted)
ordering are known, we can calculate A, B and C using the five
measured oscillation parameters in Table~\ref{tab:data}. We scan the
$\delta$ and $m_1$ ($m_3$) space to find the allowed regions, which
are defined by the condition $C^2 < A^2 + B^2$
(see Fig.~\ref{fg:1ANH} for the normal ordering and the upper panel of
Fig.~\ref{fg:1AIH} for the inverted ordering). We show allowed regions
corresponding to the best-fit parameters, and those allowed at
$2\sigma$. We also plot iso-$\phi_2$ and iso-$|M_{ee}|$ contours using
the best-fit oscillation parameters. 
We only show the contours for the plus sign of $\phi$ in
Eq.~(\ref{eq:phi}) because changing $\delta$ to $360^\circ -\delta$ yields
the same contours for the minus solution. 

The allowed regions can be further constrained using leptogenesis. We
assume the lightest right-handed neutrino has mass between $10^{12}$ GeV and
$10^{13}$ GeV and is much lighter than the others so that we can use
Eq.~(\ref{eq:etaB}). We also require $\tilde m_1 \ge 0.01$~eV.
From Eqs.~(\ref{eq:meff1}) and
(\ref{eq:epsilon1}), we see that the baryon asymmetry depends on
the sign choices of $\phi$ in Eq.~(\ref{eq:phi}) but not
on the sign choices in Eq.~(\ref{eq:solution}), because
different choices of signs in Eq.~(\ref{eq:solution}) change the
signs of all parameters in one row of the $Y$ matrix, which yield the
same baryon asymmetry. The baryon asymmetry also depends on the row of
$Y$ that is associated with the lightest right-handed
neutrino mass, but the order of the other two rows does not affect the
baryon asymmetry.

Since Classes 1A, 1B and 1C have different textures, the allowed regions
for successful leptogenesis are also different in these three
cases. Here we only consider Class 1A as an example. We find that
successful leptogenesis is not possible for the normal ordering.
For the inverted ordering, the allowed regions are shown in
the three lower panels of Fig.~\ref{fg:1AIH}. Although the constraints on
$\delta$ vary according to which right-handed neutrino is lightest, in no case
is the lightest left-handed neutrino allowed to be above 100~meV.

\subsection{Class 2 ($M_{13} = 0$)}

Similar to Class 1, the only condition for Class 2 is $M_{13} = 0$, which
is independent of ordering and can be written as
\be
m_1 = - {m_3 e^{i\phi_3} U_{\tau3}U_{e3}+m_2 e^{i\phi_2} U_{\tau2}U_{e2}\over U_{\tau1}U_{e1}}.
\ee
After taking the absolute square, then as in Class 1 this may be put
in the form of Eq.~(\ref{eq:condition}), with A, B and C as in
Table~\ref{tab:ABC}.

\begin{figure}
\centering
\includegraphics[width=6.0in]
{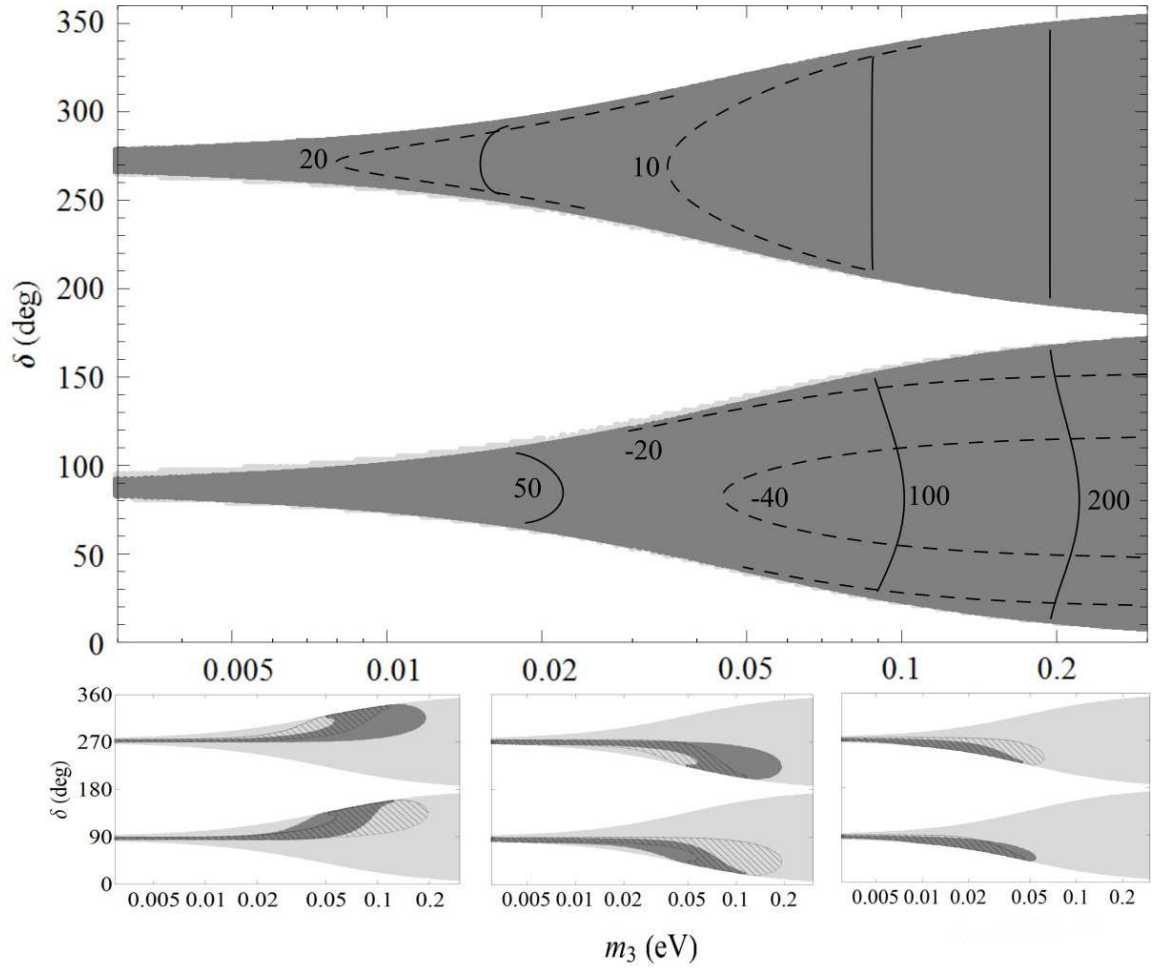}
\caption{Same as Fig.~\ref{fg:1AIH}, except for Class 2 and 2A and the
inverted ordering.}
\label{fg:2AIH}
\end{figure}

\begin{figure}
\centering
\includegraphics[width=6.0in]
{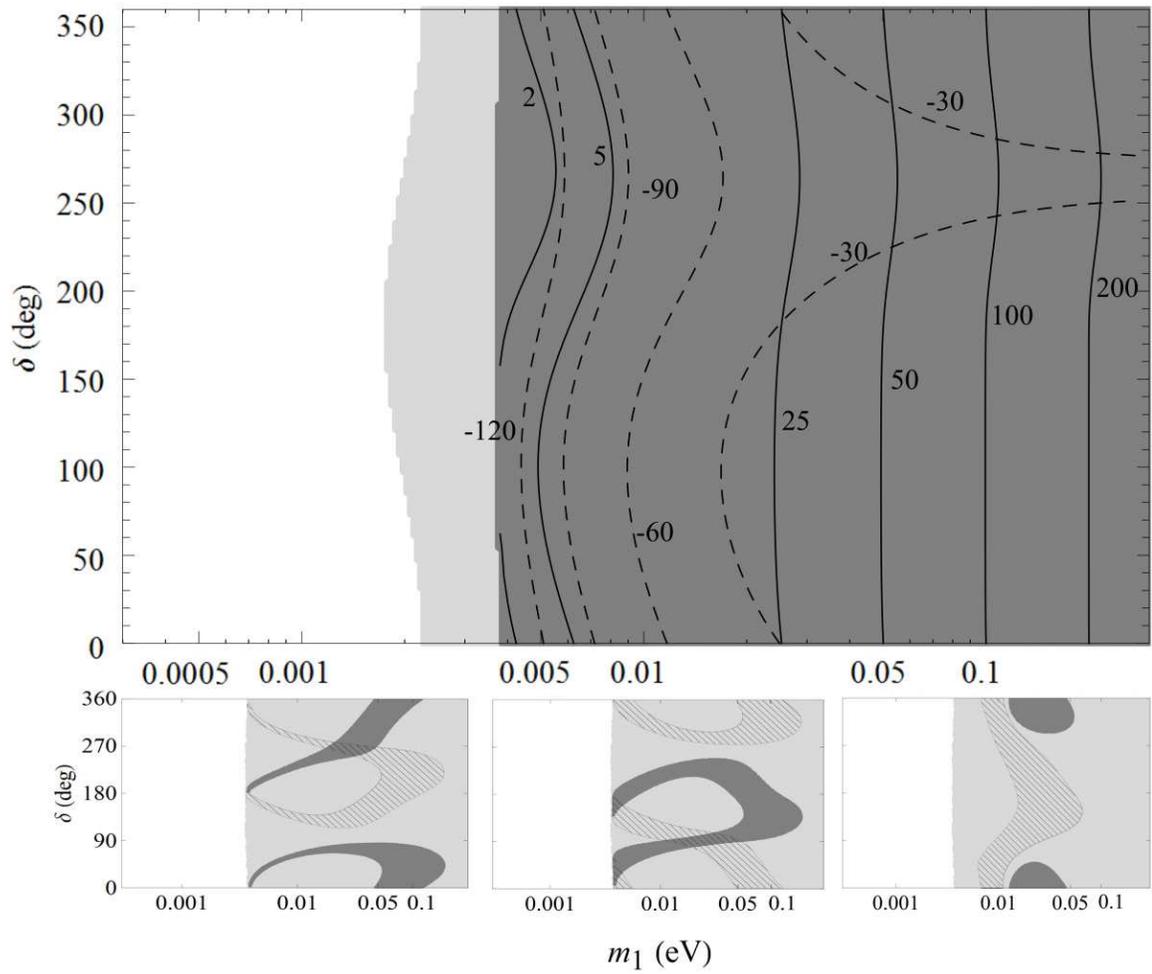}
\caption{Same as Fig.~\ref{fg:1AIH}, except for Class 3 and 3A and the
normal ordering.}
\label{fg:3ANH}
\end{figure}

The solutions for $\phi_2$ and $\phi_3$ then proceed as in Class~1.
The allowed regions for the inverted ordering are shown in
Fig.~\ref{fg:2AIH}, along with iso-$\phi_2$ and iso-$|M_{ee}|$
contours.  We see that the solution found in
Ref.~\cite{Harigaya:2012bw} with $\lambda_{1e}=\lambda_{2\tau}=0$ or
$\lambda_{1\tau}=\lambda_{2e}=0$ for the inverted ordering is a
special case of our model with $m_3=0$. Leptogenesis predictions for
Class 2A IO are also shown in Fig.~\ref{fg:2AIH}, and give an upper
bound on $m_3$ of about 200~meV. The allowed regions for the normal
ordering are similar to Fig.~\ref{fg:3ANH}.

\subsection{Class 3 ($M_{12} = 0$)}

The only condition for Class 3 is $M_{12} = 0$, which can be written
\be
m_1 = - {m_3 e^{i\phi_3} U_{\mu3}U_{e3}+m_2 e^{i\phi_2} U_{\mu2}U_{e2}\over U_{\mu1}U_{e1}}.
\ee
This condition is the same for both mass orderings and as before this may be put in the form of
Eq.~(\ref{eq:condition}) with A, B and C as in Table~\ref{tab:ABC}.

Note that Class 3 is the same as Class 2 for any ordering with
$U_{\tau j} \to U_{\mu j}$, or $s_{23} \to -c_{23}$ and $c_{23} \to
s_{23}$, which is the same as $\delta \to \delta+180^\circ$ when
$\theta_{23} = 45^\circ$. Since $\theta_{23} \approx 45^\circ$, the
phenomenology of Classes 2 and 3 will be similar.
The allowed regions for the normal
ordering are shown in Fig.~\ref{fg:3ANH}; also shown are predictions
for $|M_{ee}|$ and $\phi_2$ and regions compatible with
leptogenesis. The allowed regions for the inverted ordering are
similar to those for Class 2 in Fig.~\ref{fg:2AIH}
with $\delta \to \delta+180^\circ$. We note that the solution found in
Ref.~\cite{Harigaya:2012bw} with $\lambda_{1e}=\lambda_{2\mu}=0$ or
$\lambda_{1\mu}=\lambda_{2e}=0$ for the inverted ordering is a
special case in our model with $m_3=0$.

\subsection{Class 4A ($(M^{-1})_{23} = 0$)}

Class 4 has no texture zeros in the mass matrix. However, the existence
of a solution for $Y$ still depends on only
one condition. Take Class 4A for example, in which case
\begin{align}
Y=\begin{bmatrix}
   a & b & c \\
   0 & d & 0  \\
   0 & 0 & e
   \end{bmatrix}\,,
\end{align}
where \(a,b,c,d,e\) are all nonzero complex numbers. Then the mass matrix becomes
\begin{align}
M=Y^TY=\begin{bmatrix}
   a^2 & ab & ac \\
   ab & b^2+d^2 & bc  \\
   ac & bc & c^2+e^2
   \end{bmatrix},
\end{align}
and we see that if \(\frac{M_{11}}{M_{12}}=\frac{M_{13}}{M_{23}}\)
is satisfied, the other variables can be directly determined. Hence the
Class 4A condition is equivalent to $M_{11} M_{23}= M_{12} M_{13}$,
which means the (2, 3) cofactor of $M$, $C_{23} = M_{12} M_{31}-
M_{11} M_{32}$, vanishes. Since $(M^{-1})_{ij}=\frac{1}{\det
  M}C_{ji}$, the condition for Class 4A is equivalent to
$(M^{-1})_{23}=0$, i.e., a texture zero in the inverse mass
matrix. Since $M^{-1}=V\text{diag}(m_1^{-1}, m_2^{-1}, m_3^{-1})V^T$,
we can write the condition as
\be
m_1^{-1} U_{\tau 1} U_{\mu1}
+ m_2^{-1}e^{i\phi_2} U_{\tau2} U_{\mu2}
+ m_3^{-1}e^{i\phi_3} U_{\tau3} U_{\mu3} = 0\,,
\ee
or
\begin{equation}
m_1^{-1} = -\frac{m_2^{-1}e^{i\phi_2} U_{\tau2} U_{\mu2}
+m_3^{-1}e^{i\phi_3} U_{\tau3} U_{\mu3}}{U_{\tau 1} U_{\mu1}}.
\end{equation}

\begin{figure}
\centering
\includegraphics[width=6.0in]
{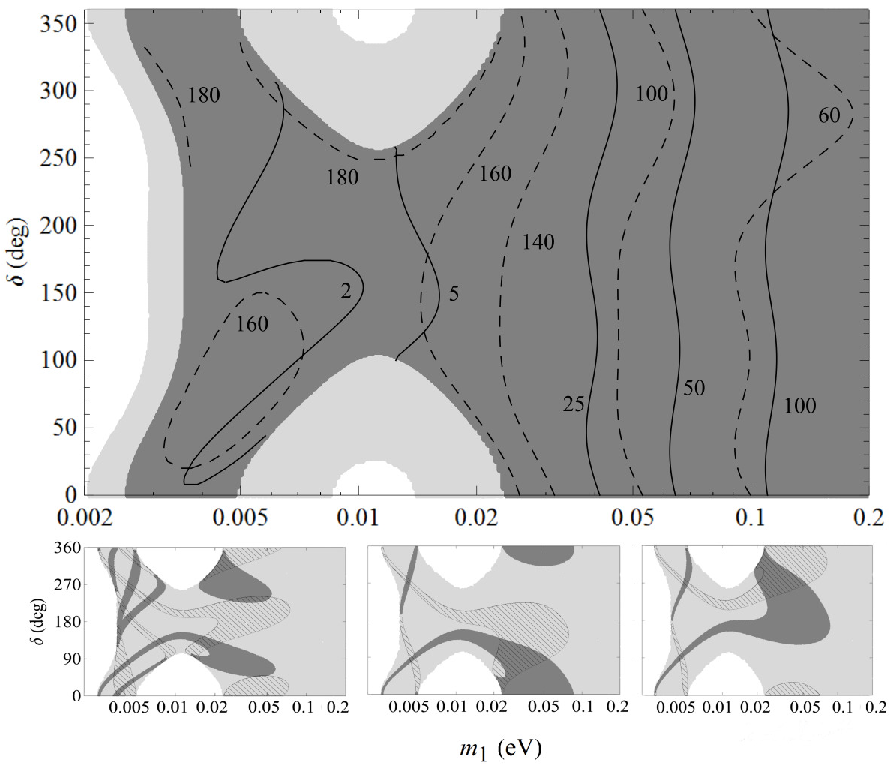}
\caption{Same as Fig.~\ref{fg:1AIH}, except for Class 4A and the
normal ordering.}
\label{fg:4ANH}
\end{figure}

The allowed regions for the normal ordering
are shown in Fig.~\ref{fg:4ANH}. The allowed regions
for the inverted ordering are similar to Fig.~\ref{fg:1ANH}
(Class 1 NO), and the iso-$\phi_2$ and iso-$|M_{ee}|$ contours are similar
with $\phi_2 \to -\phi_2$. The similarity of an IO scenario with an NO
one may seem unusual, but can be understood by looking at the form of
the $A$, $B$, and $C$ coefficients in Table~2; multiplying the
coefficients for Class 4A IO by $m_2 m_3$, and dividing the coefficients
for Class 1 NO by $m_2 m_3$, we see that $A$ and $B$ are the same for
the two cases. For the $C$ coefficient, the dominant term in each case is
the third one, proportional to $|U_{\mu3}|^2|U_{\tau3}|^2$ times the
ratio of a larger mass to a smaller one.

The comparison of Class 4A NO with Class 1 IO is more nuanced: when the
lightest mass ($m_1$ for NO, $m_3$ for IO) is very small, the first
two terms in the $C$ coefficient have similar size for Class 1 IO, but
only the first term is dominant for Class 4A NO. However, when the lightest
mass is not so small, such that $m_1 \approx m_2$ in the NO, then the
same terms in the $C$ coefficient are dominant. Thus for higher values
of the lightest mass, although not necessarily so large that all three
masses are quasi-degenerate, Classes 4A NO and 1 IO give similar
predictions.  This can be seen by comparing Figs.~5 and 2: although
the allowed regions and contours are quite different when the lightest
mass is below 20~meV, note the similarity of the $|M_{ee}| = 100$~meV
and $\phi_2 = 60^\circ$ contours.

\subsection{Class 4B ($(M^{-1})_{13} = 0$)}

Similar to Class 4A, the condition for Class 4B is $(M^{-1})_{13}=0$,
which can be written as
\begin{equation}
m_1^{-1} = -\frac{m_2^{-1}e^{i\phi_2} U_{\tau2} U_{e2}
+m_3^{-1}e^{i\phi_3} U_{\tau3} U_{e3}}{U_{\tau 1} U_{e1}}.
\end{equation}
This condition is the same for both mass orderings, and the analysis
follows as in previous classes, with A, B and C given in
Table~\ref{tab:ABC}.

\begin{figure}
\centering
\includegraphics[width=6.0in]
{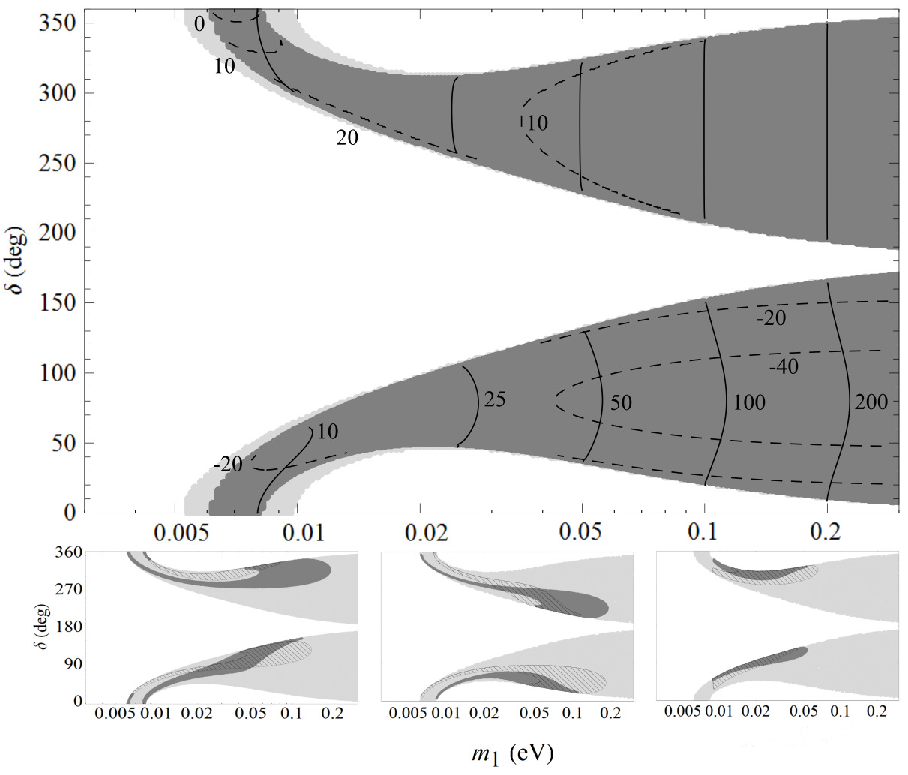}
\caption{Same as Fig.~\ref{fg:1AIH}, except for Class 4B and the
normal ordering.}
\label{fg:4BNH}
\end{figure}

\begin{figure}
\centering
\includegraphics[width=6.0in]
{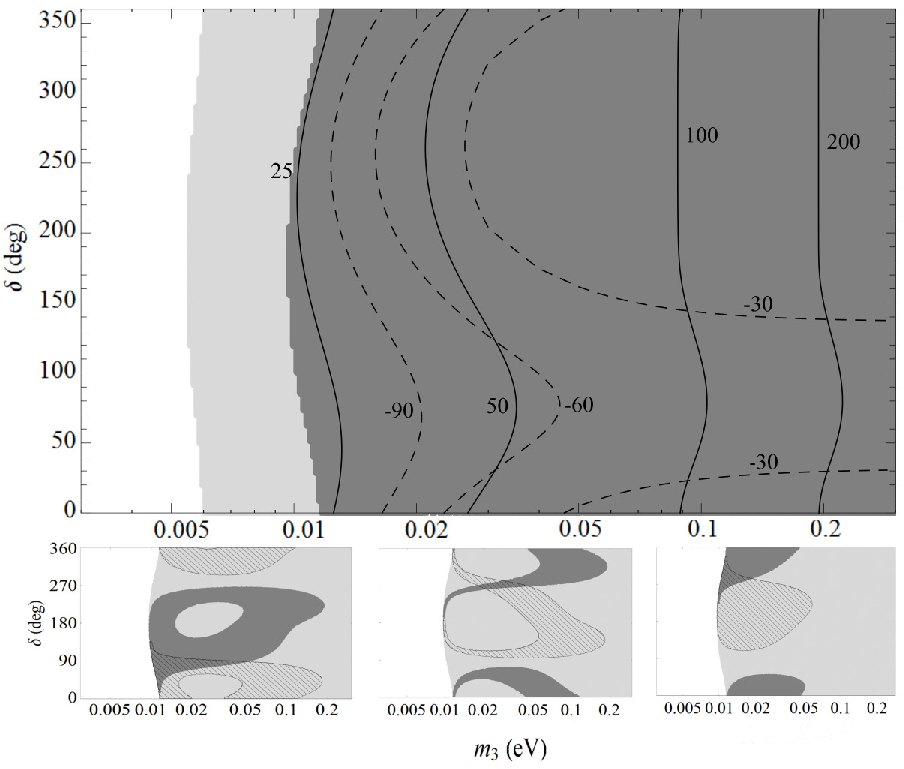}
\caption{Same as Fig.~\ref{fg:1AIH}, except for Class 4B and the
inverted ordering.}
\label{fg:4BIH}
\end{figure}

The allowed regions for the normal (inverted) ordering
are shown in Fig.~\ref{fg:4BNH} (Fig.~\ref{fg:4BIH}).
The inverted ordering for this case is also
similar to Class 2 NO: multiplying the $A$, $B$, and $C$
coefficients by $m_2 m_3$ for Class 4B IO and dividing them by $m_2
m_3$ for Class 2 NO, $A$ and $B$ are identical for the two cases, and
the dominant terms in $C$ are also the same.  As was true in the
previous section, the reverse correspondence between Class 4B NO and
Class 2 IO exists only for larger values of the lightest mass (see
Figs. 6 and 3 when the lightest mass is above 50~meV).

\subsection{Class 4C ($(M^{-1})_{12} = 0$)}

Similar to Class 4A, the condition for Class 4C is $(M^{-1})_{12}=0$,
which can be written as
\begin{equation}
m_1^{-1} = -\frac{m_2^{-1}e^{i\phi_2} U_{\mu 2} U_{e 2}+m_3^{-1}e^{i\phi_3} U_{\mu3} U_{e3}}{U_{\mu 1} U_{e1}}.
\end{equation}
The corresponding values of A, B and C in Eq.~(\ref{eq:condition}) are given in
Table~\ref{tab:ABC}.

Note that Class 4C is the same as Class 4B with $U_{\tau j} \to U_{\mu
  j}$, or $s_{23} \to -c_{23}$ and $c_{23} \to s_{23}$. As noted in
Sec.~3.3, this transformation is equivalent to $\delta \to
\delta+180^\circ$ when $\theta_{23} = 45^\circ$. Therefore the allowed
regions of Class 4C are similar to Class 4B in Fig.~\ref{fg:4BNH} for
the normal ordering and in Fig.~\ref{fg:4BIH} for the inverted
ordering with $\delta \to \delta+180^\circ$.

The inverted ordering for this case is also similar to Class 3 NO, as
can be seen by examining the $A$, $B$ and $C$ coefficients, and Class
4C NO and Class 3 IO give similar results for larger values of the
lightest mass. Thus there is a general pattern that the texture zero NO
and corresponding cofactor zero IO have similar predictions, and texture
zero IO and corresponding cofactor zero NO have similar predictions
when the lightest mass is not too small.

\section{Conclusions}

We extended the most economical type I seesaw model to include three
right-handed neutrinos. The simplest cases that fit the data have four
texture zeros in the Yukawa couplings that connect the left-handed and
right-handed neutrinos.  These are equivalent to a single texture or
cofactor zero for an off-diagonal element of the light neutrino mass
matrix $M$. The cofactor zero condition is itself equivalent to a
texture zero in $M^{-1}$. We used the latest experimental data to
obtain the allowed regions for the lightest neutrino mass and Dirac CP
phase $\delta$, which can be measured in future neutrino
experiments. We also used leptogenesis to further constrain the
allowed regions; in general there is an upper bound on the lightest
neutrino mass of about 100-200~meV for a single-flavored leptogenesis
scenario. Figures~2 to 7 show that in any given model, not all
values of $\delta$ are consistent with the leptogenesis
predictions. Therefore a precise measurement of $\delta$ will reduce
the number of viable models.

\begin{figure}
\centering
\includegraphics[width=6.0in]
{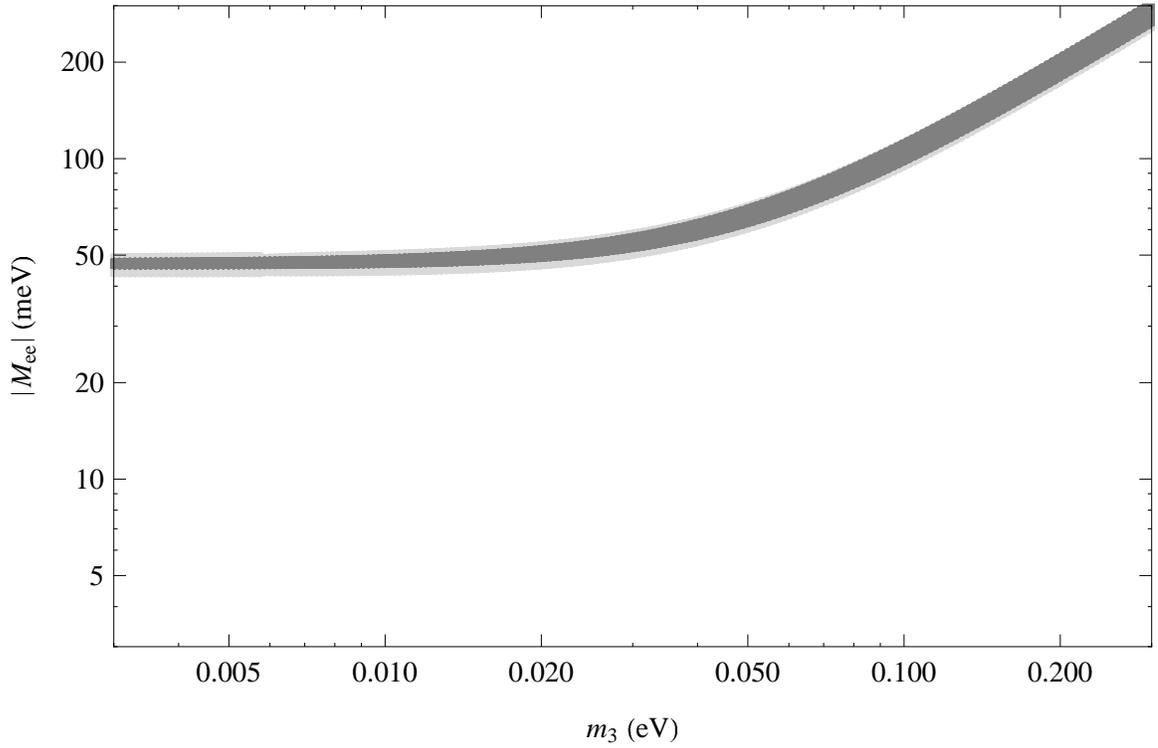}
\caption{The allowed regions in the $(|M_{ee}|,m_3)$ plane for Class 2 and
the inverted ordering. The dark shaded regions correspond to the best-fit
parameters of the oscillation parameters, while the light shaded regions
are allowed at $2\sigma$.}
\label{fg:2AIH_Mee}
\end{figure}

\begin{figure}
\centering
\includegraphics[width=6.0in]
{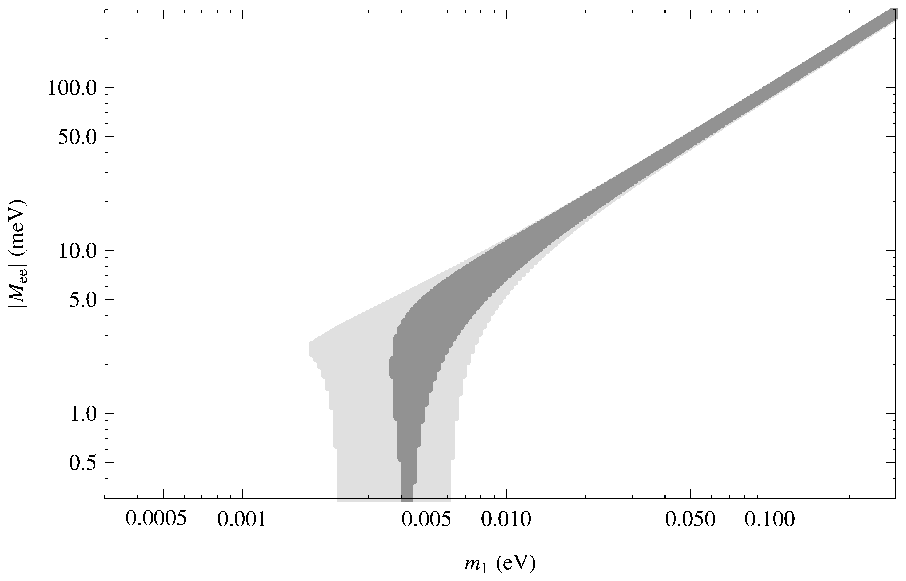}
\caption{Same as Fig.~\ref{fg:2AIH_Mee}, except for Class 3 and the
normal ordering.}
\label{fg:3ANH_Mee}
\end{figure}

\begin{figure}
\centering
\includegraphics[width=6.0in]
{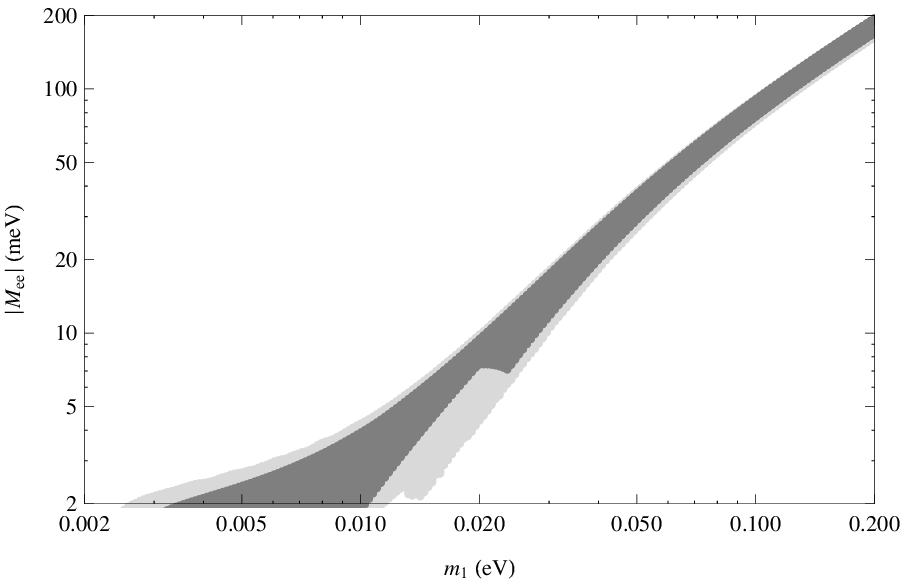}
\caption{Same as Fig.~\ref{fg:2AIH_Mee}, except for Class 4A and the
normal ordering.}
\label{fg:4ANH_Mee}
\end{figure}

\begin{figure}
\centering
\includegraphics[width=6.0in]
{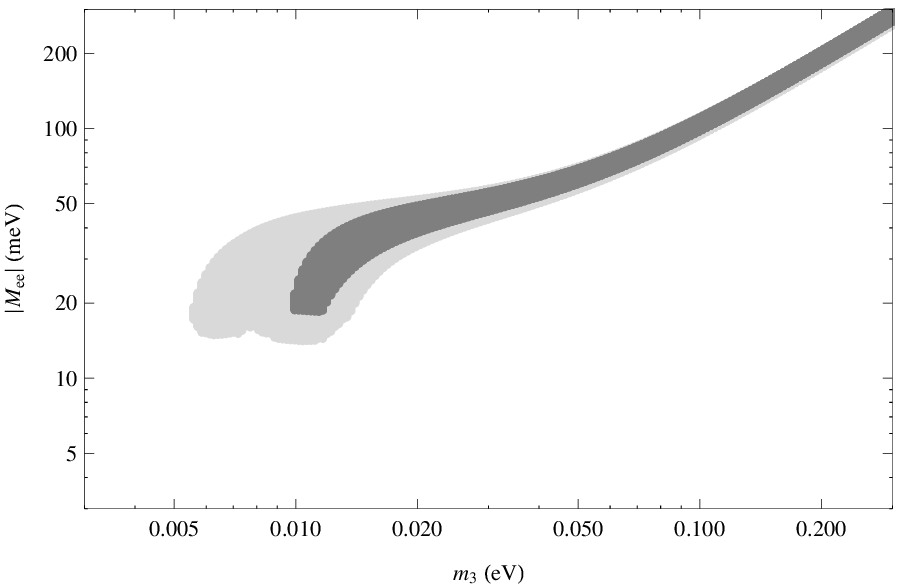}
\caption{Same as Fig.~\ref{fg:2AIH_Mee}, except for Class 4B and the
inverted ordering.}
\label{fg:4BIH_Mee}
\end{figure}

Once the lightest neutrino mass and Dirac CP phase are determined,
these models make definite predictions for neutrinoless double beta
decay. From the iso-$|M_{ee}|$ contours in Figs. 1-7 we see that $|M_{ee}|$ is generally proportional to the lightest mass. This behavior is clearly evident for the quasi-degenerate spectrum. However, for some classes, $|M_{ee}|$ is strongly dependent on the Dirac CP phase $\delta$ and is not determined by the lightest mass alone. We plot the allowed regions in the $(|M_{ee}|,\rm{lightest\ mass})$ plane for these classes separately; see Figs.~\ref{fg:2AIH_Mee},~\ref{fg:3ANH_Mee},~\ref{fg:4ANH_Mee} and~\ref{fg:4BIH_Mee}.

\begin{table}[t]
\caption{The minimum values of $|M_{ee}|$ (in $10^{-3}$ eV) in each class
for the best-fit oscillation parameters, and the $2\sigma$ lower bounds.}
\begin{center}
\begin{tabular}{|c|c|c|c|c|}\hline
\multirow{2}{*}{Class}&\multicolumn{2}{|c|}{Best-fit}&\multicolumn{2}{|c|}{$2\sigma$ lower bound}\\\cline{2-5}
                      &NO&IO&NO&IO\\\hline
1&142.5&19.0&129.8&15.4\\\hline
2&1.4&46.9&0.3&44.8\\\hline
3&0.0&47.4&0.0&45.2\\\hline
4A&0.0&150.3&0.0&138.3\\\hline
4B&6.1&18.4&5.7&15.1\\\hline
4C&8.4&18.2&7.5&14.9\\\hline

\end{tabular}
\end{center}
\end{table}

The minimum value of $|M_{ee}|$ for the best-fit oscillation
parameters and the $2\sigma$ lower bounds are shown in Table~3. We
find that the Class 1 NO and Class 4A IO have a minimum value for
$|M_{ee}|$ of around 150 (130)~meV for the best-fit parameters
(at 2$\sigma$), and could therefore be excluded by the
$0\nu\beta\beta$ decay experiments that are currently running; Classes
2 IO and 3 IO have a minimum $|M_{ee}|$ of around 50~meV and can be
definitively tested by experiments under construction. For Classes 3
NO and 4A NO the current lower bound on $|M_{ee}|$ is zero, and given
the current measurements of the oscillation parameters,
$0\nu\beta\beta$ decay will not constrain them. The remaining models
have a minimum $|M_{ee}|$ in the range $1-20$~meV and can only be
completely probed by significant improvements in the sensitivity of
$0\nu\beta\beta$ experiments. The sum of light neutrino masses can
also be used to provide additional discrimination among these
models. However, there is a general pattern that a texture zero NO and
the corresponding cofactor zero IO have similar predictions, and a
texture zero IO and the corresponding cofactor zero NO have similar
predictions when the lightest mass is not too small. Therefore it may
be difficult to uniquely specify the model from data, although
experiments designed to determine the mass ordering could resolve
this ambiguity.

Since the models studied in this paper are equivalent to a single
texture or cofactor zero for an off-diagonal element of $M_\nu$, one
might also consider examining models with a single texture or cofactor
zero for a {\it diagonal} element of $M_\nu$. Although not obtainable
from texture zeros in the Yukawa couplings in a type I seesaw model,
such models also have seven parameters in the light neutrino mass
matrix and can be analyzed in a similar fashion. We will study the
phenomenology of and possible motivation for these models in a
subsequent paper.

\section*{Acknowledgments}

JL and KW thank the University of Kansas for its hospitality during the
completion of this work. This research was supported by the
U.S. Department of Energy under Grant Nos. DE-FG02-01ER41155 and
DE-FG02-04ER41308.

\newpage

\end{document}